\documentclass{caosp305}
\usepackage{graphicx}
\usepackage{natbib}
\articleNo{IBWS01}
\pubyear{2017}
\volume{47}
\volnumber{2}
\firstpage{1}
\received{\today}
\accepted{\today}

\def\BibTeX{{\rm B\kern-.05em{\sc i\kern-.025em b}\kern-.08em
             T\kern-.1667em\lower.7ex\hbox{E}\kern-.125emX}}

\begin{document}

%

\hauthor{Dorottya\,Sz\'ecsi}
\title{How may short-duration GRBs form? \\ A review of progenitor theories}


\author{
        Dorottya\,Sz\'ecsi 
       }

\institute{
           \ondrejov \\ \email{szecsi@asu.cas.cz}
          }

\date{April 30, 2017}

\maketitle

\begin{abstract}
The origin of gamma-ray bursts (GRBs) is still a fascinating field of research nowadays. While we have been collecting more and more observationally constrained properties of GRB-physics, new theoretical results on the progenitor evolution (be it stellar or compact object) have also emerged. I review some of the most promising progenitor theories for forming a short-duration GRB. A special emphasis is put on the hypothetical case of forming a short-duration GRB through the double black hole merger scenario -- in which case we may expect to observe a gravitational wave emission too. The chemically homogeneous channel for forming a black hole binary is discussed, and the stellar progenitors (so called TWUIN stars) are introduced. The birth place of these short-duration GRBs with a gravitational wave counterpart may be low-metallicity, starforming dwarf galaxies.
\end{abstract}

%
\section{Introduction}\label{sec:intro}

At least two main categories of GRBs have been identified so far \citep{1993ApJ...413L.101K,1996ApJ...471..915K,2005Natur.437..859H,2014ARA&A..52...43B,2016Ap&SS.361..155H}. These two categories are the short-duration, spectrally hard GRBs (SGRBs) and the long-duration, spectrally soft GRBs (LGRBs). They are, based on both their temporal and spectral properties, probably released by two different astrophysical sources \citep{2004IJMPA..19.2385Z}. SGRBs typically last for several tens of milliseconds. LGRBs typically last for dozens of seconds. 

It is possible that there exist another category in between \citep[the so-called intermediate-duration GRBs,][]{2009Ap&SS.323...83H,2016Ap&SS.361..155H}, or that ultra-long GRBs should be categorized separately from the normally long ones \citep{2014ApJ...781...13L}.

In this review, I focus on progenitor models proposed for SGRBs. The simplistic picture is that while LGRBs are produced by massive stars at collapse, SGRBs are produced by the merger of two compact objects. The latter is especially interesting in the context of gravitational wave detections. Therefore, my review pays special attention to hypotheses explaining a double black hole merger accompanied by electromagnetic radiation.

Sect.~\ref{sec:NS} reviews the most recent results of those SGRB progenitor theories that involve a neutron star. Sect.~\ref{sec:BH} discusses possible scenarios for forming a SGRB from two merging black holes, which therefore may be accompanied by an observable gravitational wave emission. Sect.~\ref{sec:Discus} concludes the review, speculating about the possible host environment of SGRBs with detectable gravitational wave counterparts. 


\section{Mergers involving neutron stars}\label{sec:NS}

\subsection{NS+NS mergers}

SGRBs have typical duration values of milliseconds up to two seconds \citep{2014ARA&A..52...43B}. Thus, the progenitor model should have a dynamical timescale of milliseconds to seconds, too. Such a progenitor was suggested by \citet{Blinnikov:1984} and \citet{1989Natur.340..126E} who both proposed that the merger of two neutron stars (NS) may be responsible. 

\begin{figure}[htbp]
	\centering
	\includegraphics[width=\linewidth]{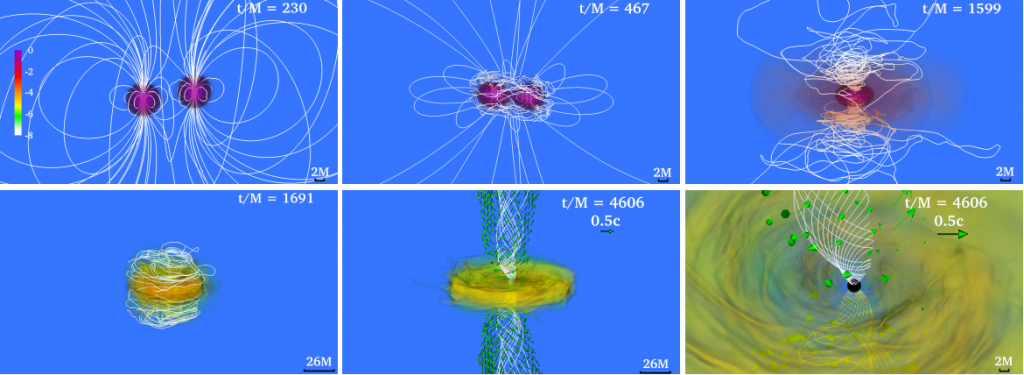}
	\caption{Snapshots of the magneto-hydrodynamic simulation of \citet{2016ApJ...824L...6R} showing the merger of two neutron stars. The coloring represents the density of the barionic material, the while lines are magnetic field lines. During the merger, a hypermassive neutron star forms (top right panel) and soon falls into a black hole (bottom right panel). Note the accretion disc and the axial jets on the bottom middle panel. The lifetime of the accretion disc, which corresponds to the lifetime of the jet, is found to be 0.1~s. It is worth to visit the website of the project where one can follow the same system's evolution on video. 
		Link for video: \textsf{http://aasnova.org/2016/06/10/jets-from-merging-neutron-stars/} 
		\textsl{Credit: \citet{2016ApJ...824L...6R}.}
	}
	\label{fig:1}
\end{figure}

Recently, \citet{2016ApJ...824L...6R} created magneto-hydrodynamic simulations of NS-NS mergers. Fig.~\ref{fig:1} shows snapshots of their results. When the hypermassive NS forms as the merger product, it sheds a significant amount of barionic material which creates an accretion disk. This disk is instrumental in the formation of the SGRB, as GRB-emission requires the presence of relativistic jets. The lifetime of the accretion disc, which corresponds to the lifetime of the jet and thus to the duration of the GRB, is found to be 0.1~s in their simulation.

\subsection{BH+NS mergers}

Another possibility for SGRB progenitor was suggested by \citet{1992ApJ...395L..83N}, namely the merger of a neutron star and a black hole (BH). This possibility was also recently simulated by \citet{2015ApJ...806L..14P}, who showed that using reasonable initial parameters for the magnetic field, it is possible to form an accretion disc around the merger product from the material shed by the NS, and that jets emerge from the system. Thus, the merger of a BH and a NS is also a promising progenitor model for SGRBs. 

Both a NS+NS merger and a BH+NS merger should, of course, emit gravitational wave radiation. However, it is still below the detection limit of our most state-of-the-art gravitational wave detector, aLIGO, to observe such an event \citep{2012MNRAS.425.2668C}.


\section{The merger of two black holes}\label{sec:BH}

As mentioned above, the prerequisite for forming a GRB is the presence of an accretion disc around the central object. In the case of a NS+BH or a NS+NS merger, the accretion disc is created from the barionic material shed by the NS or NSs. In the case of a BH+BH merger, however, there is normally no material to be shed. Therefore, the possibility of SGRB coming from BH+BH merger used to be rarely discussed until recently. 

The recent detection of gravitational wave (GW) emissions \citep{2016PhRvL.116x1103A} and the Fermi detection of a (moderately probable) electromagnetic counterpart \citep{2016ApJ...826L...6C} suddenly changed the direction of research, and many people started to theorize about possible channels through which even a BH+BH merger may produce a SGRB. Although these hypotheses of a BH+BH merger with an SGRB do not reach the level of sophistication NS+NS or NS+BH models possess, the simultaneous detection of a future GW emission and a SGRB holds huge scientific potential. Therefore, it is interesting to discuss this possibility from a theoretical point of view. 

\subsection{Chemically homogeneously evolving massive binaries, and a dead disc: GW+SGRB}

One interesting hypothesis was presented by \citet{2016ApJ...821L..18P}, which is built on the work of \citet{2016A&A...588A..50M}. The latter presented massive binary evolutionary models at low-metallicity, and showed that both stars in these models avoid the supergiant phase due to chemically homogeneous evolution. Avoiding the supergiant phase means they also avoid the common envelope phase -- a phase which is currently undergoing serious investigation \citep{2016Natur.534..512B,2016A&A...596A..58K} but is still weighted with many uncertainties in the case of massive stars. But when both massive stars evolve chemically homogeneously, they do not have a common envelope phase. Rather, they reach a contact phase during which they exchange mass. As a result, the companions have almost equal masses at the end of their main-sequence evolution. After the main-sequence phase, they shrink further and become fast rotating helium stars. Since the mass ratio is close to one, their explosions will happen soon after one another. As for the explosion, a supernova of type~Ib/c accompanied by a LGRB is expected in both cases. The LGRBs may be observed only if the jets are along the line-of-sight.

After the explosions, the remnants are two black holes orbiting around each other. They are losing orbital energy via gravitational wave radiation, slowly spiraling in. \citet{2016A&A...588A..50M} showed that their spiral-in may be well within the Hubble time. When they merge, they emit a well-defined signal of gravitational wave radiation, which can be searched for in the aLIGO data. 

\citet{2016ApJ...821L..18P} suggested that since these stars rotate very fast in the moment of their core-collapse, it is possible that -- in case of a weak supernova explosion -- at least one of them keeps a disc. They hypothesize that accretion of this disc stops soon after the core-collapse. The disc then cools down, suppressing the magnetorotational instability and hence the viscosity, and becomes inactive for a long time. This way, the disk may survive the slow spiral-in of the BHs. But when the companion BH reaches the outer rim of the disk, the disk becomes active again. Thus, we have a situation when a BH with an active accretion disc is merging with another BH. This opens the possibility for a SGRB while the BH+BH merger is producing detectable GW emission in the same time. \citet{2016ApJ...821L..18P} estimated a GRB timescale of 0.005~s.

\begin{figure}[htbp]
	\centering
	\includegraphics[width=0.7\linewidth]{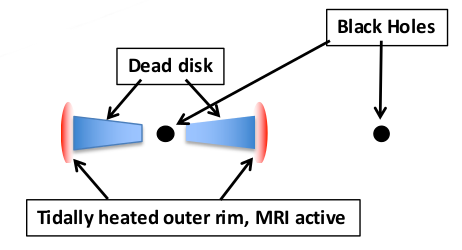}
	\caption{Set-up of the hypothetical scenario forming a double black hole merger together with an electromagnetic counterpart. Once the two black holes come close enough, the disc is re-heated and accretion starts again.
		\textsl{Credit: \citet{2016ApJ...821L..18P}.}
	}
	\label{fig:2}
\end{figure}

\subsection{But what kind of stars are these?}

The double black hole system was formed from a special type of massive star binary. This binary is on a very close orbit and is, therefore, initially synchronized (i.e. the orbital period is the same as the spin period of the companions). These stars not only orbit around each other fast but spin fast, too \citep{2009A&A...497..243D,2016MNRAS.460.3545D,2016MNRAS.458.2634M}.

At low-metallcity ($\sim$1/10 of solar and below) fast rotating massive stars evolve chemically homogeneously \citep{2006A&A...460..199Y,2011A&A...530A.115B,2015A&A...573A..71K,2015A&A...581A..15S}. This means that they do not develop a distinct core-envelope structure as typical massive stars do. Instead, they are homogeneously mixed. Their surface composition reflects the nuclear burning of the core. They do not become supergiants, but stay rather small (a few tens of solar radii) during all their evolution. At the end of their main-sequence phase, they consist almost entirely of helium. 

Many such stellar models were created and analysed by \citet{2015A&A...581A..15S}. They showed that although some of the surface properties (e.g. temperature, composition) are similar to those of Wolf--Rayet (WR) stars, these chemically homogeneously evolving objects have \textsl{weak} stellar winds. Therefore, they are not WR stars in the classical sense: WR stars are observed to have strong, optically thick winds. \citet{2015A&A...581A..15S} called these objects Transparent Wind Ultraviolet Intense (TWUIN) stars. TWUIN stars are the result of chemically homogenenous evolution. 

\begin{figure}[htbp]
	\centering
	\includegraphics[width=0.9\linewidth]{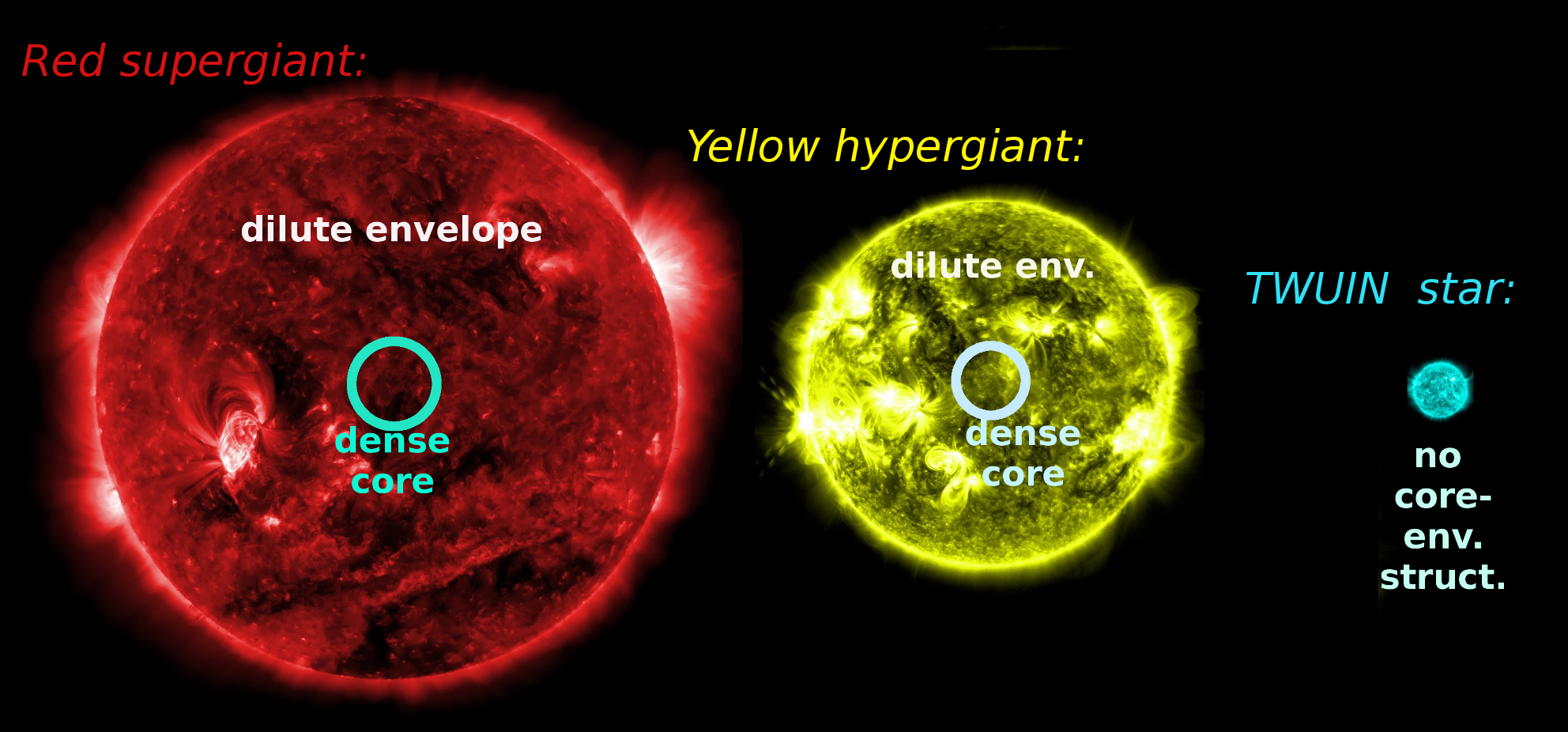}
	\caption{Schematic structure of some massive stars. While supergiants and hypergiants have a distinct core-envelope structure, TWUIN stars are rather compact ($\sim$10-20~R$_{\mathrm{solar}}$), homogeneous, hot objects. 
	}
	\label{fig:3}
\end{figure}

This evolutionary channel of massive stars is only possible at low-metallicity. The reason is the following. In the case of massive stars with line-driven stellar winds, the mass-loss rate scales with metallicity. Since losing mass means losing also angular momentum, with a stronger mass-loss comes a faster spin-down. Thus, massive stars at high metallicity (e.g. at solar composition) typically do not rotate fast. Fast rotation, however, is needed for chemically homogeneous evolution: the process responsible for keeping the whole star homogeneous is called rotational mixing. The effectivity of rotational mixing scales with the rotational rate. Therefore, according to our most state-of-the-art stellar models, chemically homogeneous evolution can only happen at low-metallicity. 

If born in a close binary system, chemically homogeneously evolving TWUIN stars are expected to produce two consequent supernova explosions of type Ib/c (and a LGRB, as explained above). If at least one the the supernovae is weak, some fractions of the stellar material may stick on a circumstellar orbit and may cool. Then the two remnant BHs are slowly spiralling in, and when the merger happens, the re-vitalized accretion disc provides the conditions for a SGRB. In this theory, \textsl{TWUIN binaries are the stellar progenitors of an exotic explosion which emits both gravitational waves and electromagnetic radiation.} 

\subsection{Alternative hypothesis: charged BHs}

Another promising hypothesis was presented by \citet{2016ApJ...827L..31Z}. In his theory, there is no barionic material present in the BH+BH merger system. On the other hand, one of the BHs is carrying a high amount of charge. The rapidly evolving magnetic moment during the merger may first lead to a fast radio burst (FRB), and then a SGRB. \citet{2016ApJ...827L..31Z} suggest that combined with future GW detections with an electromagnetic counterpart, this theory may put constraint on the charges carried by isolate BHs.


\section{Discussion}\label{sec:Discus}

\subsection{Dwarf galaxies as birthplaces of GW+SGRB events}

TWUIN stars are theoretical predictions at low-metallicity. However, their presence in irregular, star-forming dwarf galaxies has been recently considered based on the fact that TWUIN star models predict that a huge amount of ionizing photons is emitted in the He~II continuum \citep{2015A&A...581A..15S}. Some of the dwarf galaxies are observed to have an ionized interstellar medium, but no traces of any other ionizing source \citep[such as X-ray binaries or WR stars,][]{2015ApJ...801L..28K}. Therefore, \citet{2015wrs..conf..189S} speculated that the extent of ionization in dwarf galaxies may be an indication for the existence of chemically homogeneous evolution and, therefore, for TWUIN stars. Since massive stars are almost always in binaries \citep{2012Sci...337..444S}, some of them close binaries \citep{2009A&A...497..243D}, TWUIN binaries may be common in low-metallicity dwarf galaxies. \textsl{Dwarf galaxies are therefore expected to host the event of a GW emission accompanied by a SGRB.}

\subsection{Detections and theories}

From an observational point of view, simultaneous detection of a GW and a SGRB is challenging, but well possible with our current instrumentation \citep{2016ApJ...826L...6C}. It is indeed an interesting data mining problem how to identify faint electromagnetic counterparts for any GW event \citep[][]{2013A&A...557A...8S,2016A&A...593L..10B}. 

In case such a simultaneous detection happens, it is expected that more and more theories will emerge. These new theories will either build on and expand, or contradict the theories described above. Hence, these are exciting times not only for observers but for theorists, too.

\acknowledgements
D. Sz. was supported by the Czech grant 13-10589S \v{C}R.


\bibliography{Szecsi-review}

\end{document}